\documentclass[12pt]{iopart}
\usepackage{epsfig}
\usepackage{graphicx}
\newcommand{\beq}[1]{\begin{equation}\label{#1}}
\newcommand{\eeq}{\end{equation}}
\newcommand{\bear}[1]{\begin{eqnarray}\label{#1}}
\newcommand{\ear}{\end{eqnarray}}
\newcommand{\nn}{\nonumber}

\newcommand{\fii}{\varphi}
\newcommand{\om}{\omega}

\begin{document}
\title[Friedmann universe with dust and scalar field]
{Flat Friedmann universe filled by dust and scalar field with
multiple exponential potential}

\author{V R Gavrilov\footnote[3]{To whom
correspondence should be addressed (gavr@rgs.phys.msu.su)}, V N
Melnikov and  S T Abdyrakhmanov}

\address{Centre for
Gravitation and Fundamental Metrology, VNIIMS, and Institute for
Gravitation and Cosmology, PFUR, 3-1 M.Ulyanovoy St., Moscow
117313, Russia}

\eads{\mailto{gavr@rgs.phys.msu.su},
\mailto{melnikov@rgs.phys.msu.su} and
\mailto{luxon@rgs.phys.msu.su}}

\begin{abstract}
We study a spatially flat Friedmann model
containing a pressureless perfect fluid (dust) and a scalar
field with an unbounded from below potential of the form
$V(\fii)=W_0 - V_0\sinh\left(\sqrt{3/2}\kappa\fii\right)$, where
the parameters $W_0$ and $V_0$ are arbitrary and
$\kappa=\sqrt{8\pi G_N}=M_p^{-1}$. The model is integrable and
all exact solutions describe the recollapsing universe. The
behavior of the model near both initial and final points of
evolution is analyzed. The model is consistent with the
observational parameters. We single out the exact solution with
the present-day values of acceleration parameter $q_0=0.5$
and dark matter density parameter $\Omega_{\rho 0}=0.3$
describing the evolution within the time approximately equal
to $2H_0^{-1}$.
\end{abstract}

\submitto{\it Gen. Rel. Grav.
}

 \pacs{04.20.Jb}

\maketitle
\section{Introduction}

    Scalar fields play an essential role in modern cosmology.
They are attributed to inflation models of the early universe and
the models describing the present stage of the accelerated
expansion as well. There is no unique candidate for the potential
of the minimally coupled scalar field. Typically a potential is a
sum of exponentials. Such potentials appear quite generically in a
large class of theories (multidimensional \cite{Mel1} -
\cite{Mel5}, Kaluza-Klein models , supergravity and string/M -
theories, see, for instance, \cite{FJ}, \cite{T} and references
therein) and from reconstruction and other schemes \cite{star}.

 Single exponential potential was
extensively studied within the Friedmann
model using both qualitative methods \cite{Wands} and exact
solutions \cite{Dehnen}. As to the multiple exponential potential,
it's not studied well yet.
Here we consider some multiple potential unbounded from below.
According to \cite{Linde}
a Friedmann universe involving a scalar field with an
unbounded from below potential exhibits the interesting features such
as recollapsing even if it's flat.

In this paper we study  the spatially flat Friedmann model containing a
pressureless perfect fluid (dust) and a scalar field with an
unbounded from below simplest multiple exponential potential of
the form $V(\fii)=W_0 -V_0\sinh\left(\lambda\sqrt{8\pi
G_N}\fii\right)$, where the parameters $W_0$ and $V_0$ are
arbitrary and $\lambda=\sqrt{3/2}$. For this value of $\lambda $
the model is integrable and all exact solutions describe the
recollapse of the universe within the finite time. The time as
well as the intermediate behavior of the model crucially depend on
parameters $W_0$ and $V_0$. The paper is organized as follows. In
section 2 we describe the model and obtain the equations of
motion. The equations are integrated in section 3 using methods
developed in our previous papers \cite{Mel1}-\cite{Mel3} and the
general properties of these exact solutions are studied. In
section 4 we fit the model to the observational parameters.

\section{The model and equations of motion}

We start with the spatially flat Friedmann universe filled by both a
self-interacting scalar field $\fii$ with a potential $V(\fii)$ and a
separately conserved pressureless perfect fluid (dust). The corresponding
Einstein equations read

\bear{1}
H^2  =  \frac{\kappa^2}{3}
\left[\frac{1}{2}\dot\fii^2+V(\fii)+\rho\right],\\
\label{2}
2\dot H + 3 H^2  =  -\kappa^2
\left[\frac{1}{2}\dot\fii^2-V(\fii)\right],
\ear
where $H=\dot a/a$ is the Hubble parameter, $\rho$ denotes the energy
density of dust, overdot means derivative with respect to the cosmic time
$t$, $\kappa=\sqrt{8\pi G_N}=M_p^{-1}$. Moreover, we have the equation
of motion of the scalar field
\beq{3}
\ddot\fii=-3H\dot\fii-V'(\fii)
\eeq
and the conservation equation for the perfect fluid
\beq{4}
\dot\rho=-3H\rho.
\eeq
The last equation gives immediately
\beq{5}
\rho = \rho_0 \left(\frac{a_0}{a}\right)^3,
\eeq
where the zero subindex means the present time, as usually.
In what follows we consider the scalar field potential of the form
\beq{6}
V(\fii)=W_0 - V_0\sinh\left(\sqrt{3/2}\kappa\fii\right),
\eeq
where $W_0$ and $V_0$ are arbitrary constants. As the system is
symmetrical under the transformation $\fii\to -\fii$, $V_0\to -V_0$,
without a loss of generality we consider only the case $V_0>0$.

Now we introduce new variables $x$ and $y$ by the following transformation
\beq{7}
a^3=xy,\ \kappa\fii=\sqrt{2/3}\log(y/x),\ x>0,\ y>0.
\eeq
Then the set of Eqs.(\ref{2}),(\ref{3}) results in
\bear{8}
\ddot x= (\om_1^2-\om_2^2)x - 2\om_1\om_2 y,\\
\label{9}
\ddot y= 2\om_1\om_2x  +  (\om_1^2-\om_2^2) y,
\ear
where we introduced the positive parameters $\om_1$ and $\om_2$ by
\bear{}
\om_1^2-\om_2^2=3/4\kappa^2W_0,\ 2\om_1\om_2=3/4\kappa^2V_0.\nn
\ear
Then Eq.(\ref{1}), where the presence of $\rho$ is cancelled by
Eq.(\ref{5}), takes the form of the following constraint
\beq{10}
\dot x\dot y-(\om_1^2-\om_2^2)xy-\om_1\om_2(x^2-y^2)
=3/4\kappa^2\rho_0 a_0^3.
\eeq
The set of Eqs.(\ref{8}),(\ref{9}) may be presented in the following
complex form
\beq{11}
\ddot z = \om^2 z,
\eeq
where we introduced the complex variable
\bear{}
z = x + \imath y \nn
\ear
and the complex parameter
\bear{}
\om = \om_1 + \imath \om_2.\nn
\ear
It is easy to see that Eq.(\ref{11}) implies the following complex
integral of motion
\bear{}
\dot z^2-\om^2 z^2={\rm const},\nn
\ear
The presence of this complex integral is equivalent to the existence of 2
real integrals of motion. One of them
$\Im(\dot z^2-\om^2 z^2)=-3/2\kappa^2\rho_0 a_0^3$
represents the
constraint given by Eq.(\ref{10}) and the other integral
$\Re(\dot z^2-\om^2 z^2)$ has an arbitrary value.

\section{Exact solutions and behavior of the model}

The equations of motion in form Eq.(\ref{11}) are easily integrable.
The result is
\bear{12}
z=C\rme^{\imath\alpha}
\left(
\rme^{\om_1(t-t_0)+\imath\om_2(t-t_0-\delta)}
+\rme^{-\om_1(t-t_0)-\imath\om_2(t-t_0-\delta)}
\right),
\ear
where $C>0$. The constants $t_0$ and $\delta$ are arbitrary,
$C$ and $\alpha$ obey the relation
\bear{}
\Im\left(\om^2C^2\rme^{\imath 2\alpha}\right)=
\frac{3}{8}\kappa^2\rho_0a_0^3\nn
\ear
following from Eq.(\ref{10}). Then substituting $x=\Re z$ and $y=\Im z$
into the relations given by Eq.(\ref{7}) one easily gets the explicit
expressions for the scale factor $a$ and the scalar field $\fii$.

Let us now consider the general properties of the exact
solutions by analyzing  the corresponding to
Eq.(\ref{12}) trajectories (orbits) of the moving point on the
Cartesian  $xy$ plane. It follows from the definition of the
variables $x$ and $y$ given by Eq.(\ref{7}) that the physical
segments of a trajectory belong to the angular domain with
$x>0$ and $y>0$. Hereafter we show that each segments is of
a  finite length and its end-points are attached either to one coordinate
axis or the both axes. From the physical viewpoint it means
that all solutions describe the universe evolution within a finite time
interval. Moreover, as the equations of motion are invariant with respect
to the time reflection $t\to -t$ each segment of the trajectory may be
passed in both directions. We notice that the constant factor
$\exp(\imath\alpha)$ leads to the rotation of the trajectory about the
origin $(0,0)$ through the angle $\alpha$. Further we consider only the
case when $\alpha=0$ taking into account that all remaining
trajectories may be obtained by rotation.

\begin{figure}
 \rotatebox{-90}{\epsfxsize=5.5cm \epsfbox{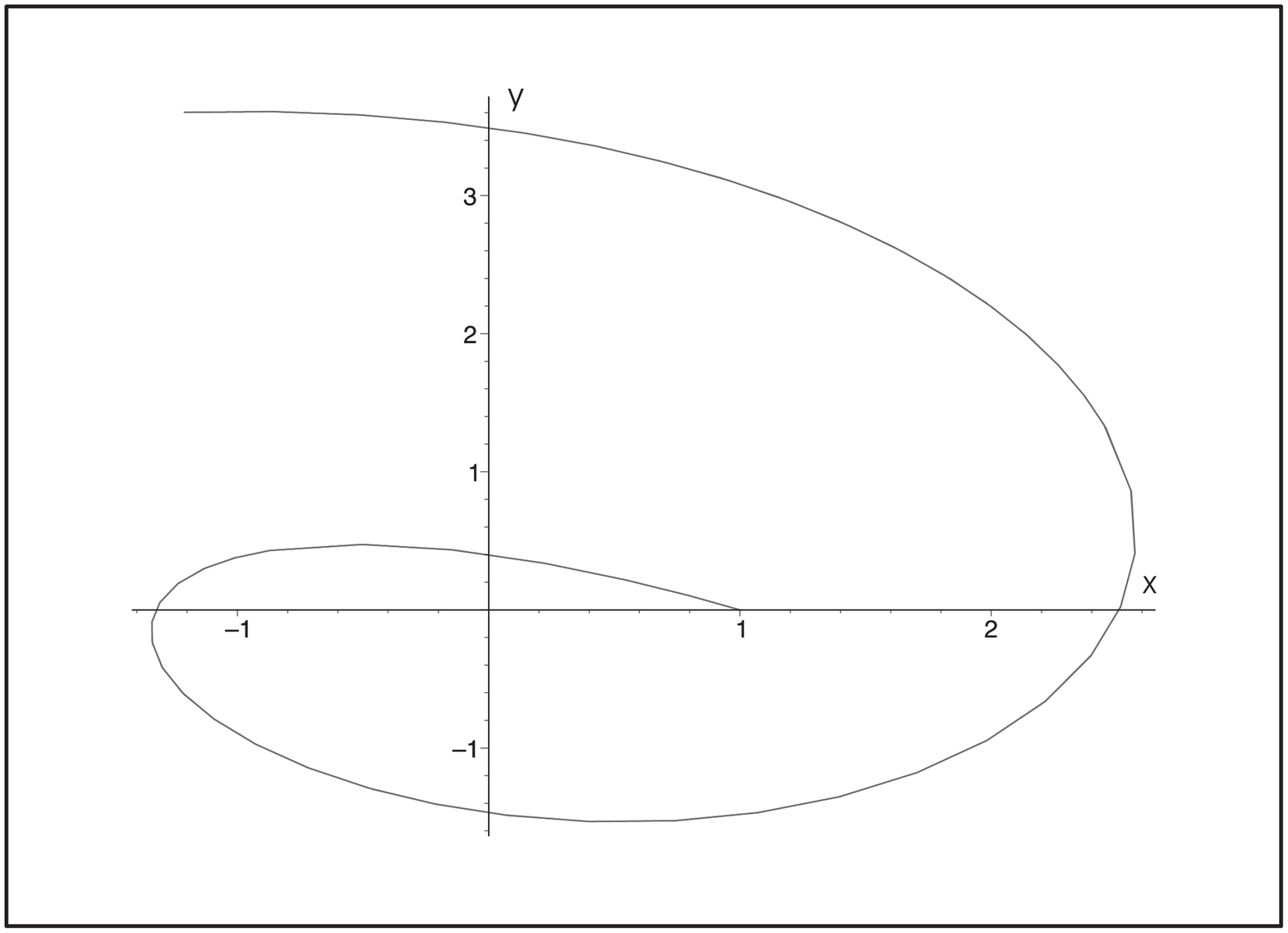}}
\hfill
 \rotatebox{-90}{\epsfxsize=5.5cm\epsfbox{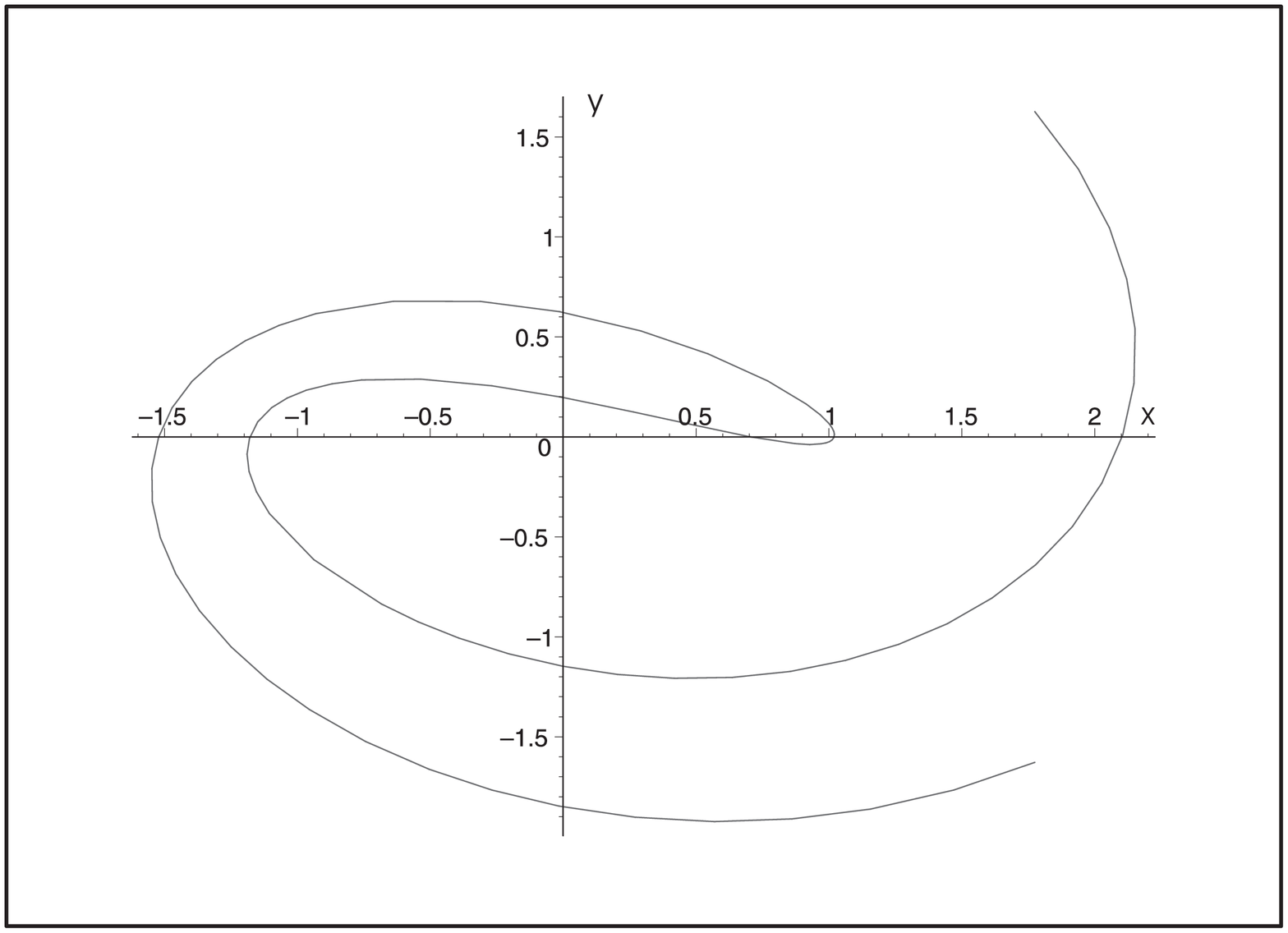}}
 \\
\parbox[c]{8cm}{\parindent=2.5cm \bf Figure 1a }
 \hfill
\parbox[c]{8cm}{\parindent=2.5cm \bf Figure 1b }
\\
\vskip2mm
 \rotatebox{-90}{\epsfxsize=5.5cm\epsfbox{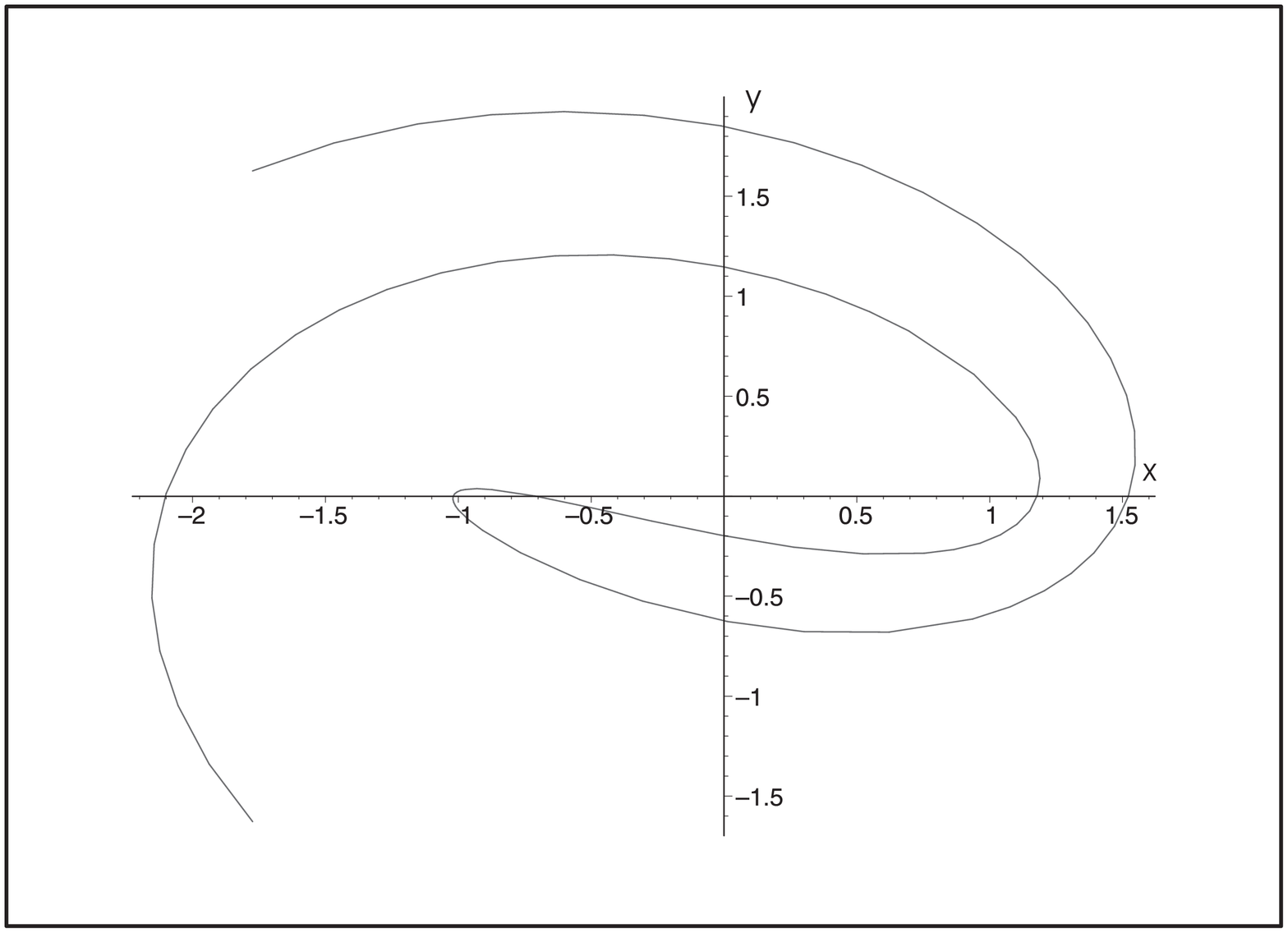}}
\hfill
\rotatebox{-90}{\epsfxsize=5.5cm\epsfbox{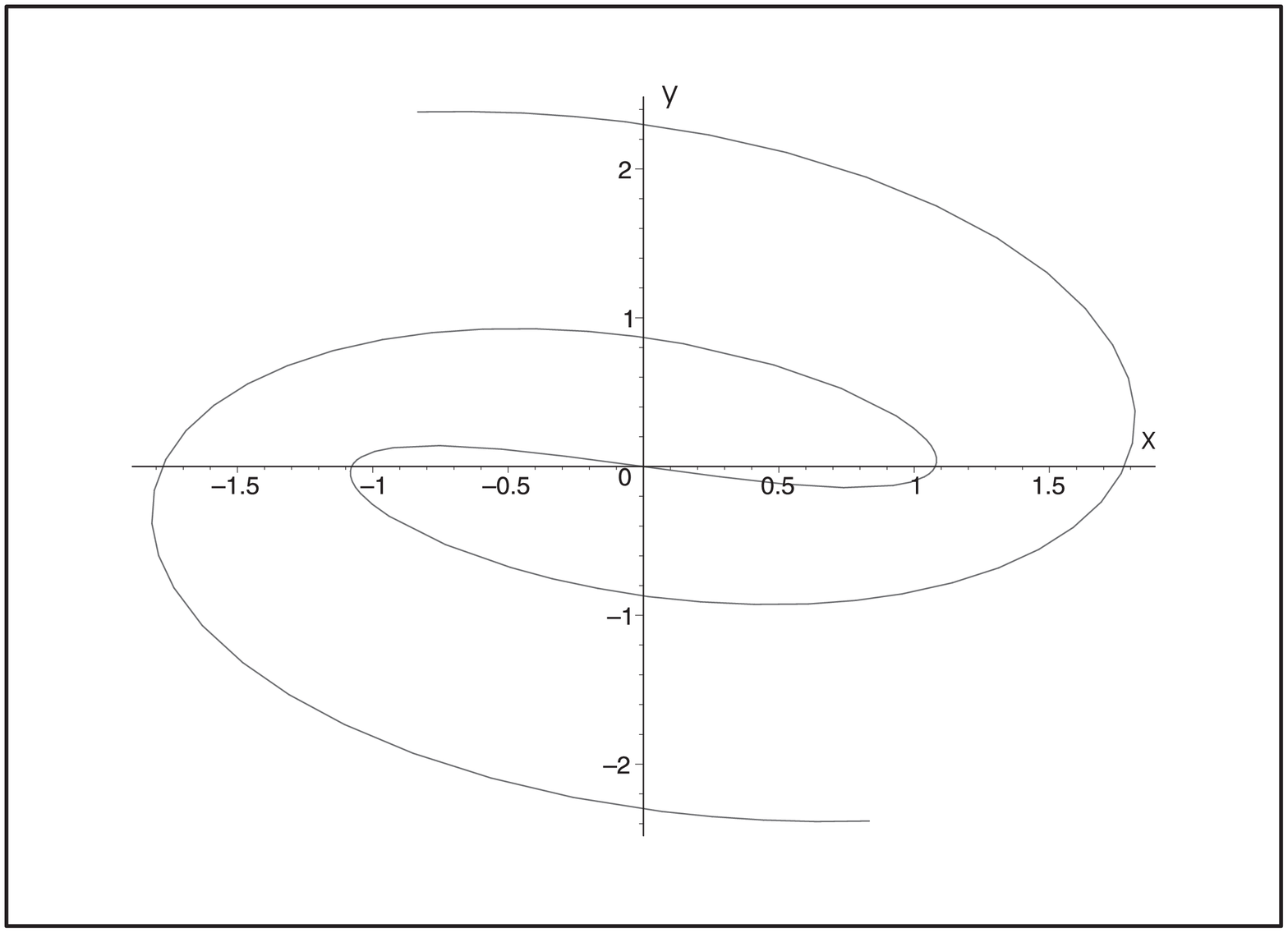}}
\\
\parbox[c]{8cm}{\parindent=2.5cm \bf Figure 1c }
\hfill
\parbox[c]{8cm}{\parindent=2.5cm \bf Figure 1d }\\
\vskip1cm
\parbox[c]{14cm}
{ \caption{The corresponding to the exact solutions
trajectories on the $xy$ plane. All remaining trajectories may be
obtain by rotation about the origin $(0,0)$. Each solution is
defined on some finite time interval which corresponds to a lying
in the angular domain with $x>0$ and $y>0$ segment of the
trajectory. End-points of such segment correspond to big bang and
big crunch.}}
\end{figure}

If the time variable $t$ is positive and large enough the last
term in Eq.(\ref{12}) is negligible ($\om_1>0$).  Then the
motion is confined to the repelling logarithmic spiral with
$x=C\exp[\om_1(t-t_0)]\cos[\om_2(t-T_0)]$ and
$y=C\exp[\om_1(t-t_0)]\sin[\om_2(t-T_0)]$, where $T_0\equiv
t_0+\delta$. As the time $t$ grows the point clockwise  rotates around
the origin $(0,0)$  and its distance from the  origin increases as
$\exp(\om_1t)$. It means that the evolution starts at some moment $t=T_0$
when the spiral intersects the axis $x$ and finishes when it further
intersects the axis $y$. The finite time of evolution is about
the value $T=\pi/(2\om_2)$.  In this case the zeros of the functions
$x(t)$ and $y(t)$ are different. Then near the initial point the expansion
of the universe can be approximately described by the power-law equation
$a(t)\sim (t-T_0)^{1/3}$. The similar equation $a(t)\sim
[T-(t-T_0)]^{1/3}$ approximately describes the collapsing near the final
point of evolution.

If $t<0$ and $|t|$ is large enough the first term in
Eq.(\ref{12}) vanishes and the last term dominates. In this case
the behavior of the model is similar with the only
difference: the corresponding spiral is attracting. It describes
the counterclockwise rotation with approaching to the origin of
coordinates $x$ and $y$.

The form of trajectories for the intermediate values of the time
variable $t$ is more complicated. To analyze the behavior one
needs to superpose both attracting and repelling spirals.
The result depends on the constant $\delta$. Evidently it's enough
to consider $\om_2\delta\in [0,\pi)$.
If $\delta=0$ we get $x=C\cosh[\om_1(t-t_0)]\cos[\om_2(t-t_0)]$
and $y=C\sinh[\om_1(t-t_0)]\sin[\om_2(t-t_0)]$. The
corresponding trajectory with $C=1$ is presented on Fig. 1a.
The moving  point passes the curve twice in both directions. It's not
difficult to prove that for $\delta\neq 0$ any trajectory has no
selfintersection points. On Figs. 1b and 1c the constant
$\om_2\delta$ ranges from $0$ to $\pi/2$ and from $\pi/2$ to $\pi$,
respectively. For $\om_2\delta=\pi/2$ the trajectory passes through
the origin center (see Fig. 1d).

This analysis shows that all exact solutions describe the
evolution from big bang to big crunch within some finite time.
Typically this time is about $T=\pi/(2\om_2)$, however it
may be arbitrarily shorter by choosing trajectories obtained after rotation.
 Each solution is defined on some
finite time interval which corresponds to a lying in the angular
domain with $x>0$ and $y>0$ segment of the trajectory. End-points
of such segment correspond to big bang and big crunch. If
end-points of the segment are attached to different coordinate
axes then  the scalar field goes from $-\infty$ to $+\infty$ or
vice versa during the evolution. If both end-points are attached
to the coordinate axis $y$, the scalar field diverges to $+\infty$
as the time tends to the initial or final value. For almost all
solutions the scale factor is proportional to the time in power
$1/3$ near the initial and the final points of evolution. But if
the end-point of the segment is in the origin $(0,0)$, the
roots of the functions $x(t)$ and $y(t)$ coincide and the scale
factor is proportional to the time in power $2/3$ near the
corresponding  point.

\section{Fitting the model to observational parameters}

After this qualitative description of the obtained exact solution
(putting the present-day value of time $t$ equal to zero) we
present the solution to the equations of motion Eq.(\ref{11}) in
the form \bear{13} z=(x_0+\imath y_0)\cosh \om t +
\frac{\dot{x}_0 + \dot{y}_0}{\om}\sinh\om t \ear with the values
$x_0\equiv x(0)$, $y_0\equiv y(0)$, $\dot{x}_0\equiv\dot{x}(0)$
and $\dot{y}_0\equiv\dot{y}(0)$ expressed in the terms of the
observational parameters: the Hubble constant $H_0$, the
present-day values $q_0$ and $\Omega_{\rho 0}$ of the
acceleration parameter $q=a\ddot{a}/\dot{a}^2$ and the density
parameter $\Omega_{\rho}=\kappa^2\rho/(3H^2)$. One easily gets
these expressions from Eqs.(\ref{1}),(\ref{2}) and the definition
of the variables $x$ and $y$ given by Eq.(\ref{7})
\bear{14}
x_0=\left( \sqrt{1+\varepsilon^2_0}+\varepsilon_0\right)^{1/2} a_0^{3/2},\\
\label{15}
y_0=\left( \sqrt{1+\varepsilon^2_0}-\varepsilon_0\right)^{1/2} a_0^{3/2},\\
\label{16}
\dot{x}_0=\frac{3}{2}x_0H_0
\left(1\mp\sqrt{\frac{1}{3}(1-q_0)-\frac{1}{2}\Omega_{\rho 0}}\right),\\
\label{17}
\dot{y}_0=\frac{3}{2}y_0H_0
\left(1\pm\sqrt{\frac{1}{3}(1-q_0)-\frac{1}{2}\Omega_{\rho 0}}\right),
\ear
where we denoted
\bear{18}
\varepsilon_0=\frac{9H_0^2}{8\om_1\om_2}
\left[ \frac{1}{3}(2+q_0)-\frac{1}{2}\Omega_{\rho 0}\right]
-\frac{\om_1^2-\om_2^2}{2\om_1\om_2}.
\ear
The upper sign in Eqs.(\ref{16}),(\ref{17}) corresponds to the positive
value $\dot{\fii}_0\equiv\dot{\fii}(0)$, the lower sign appears when
$\dot{\fii}_0<0$. We notice that the model implies the value
$(1-q_0)/3-\Omega_{\rho 0}/2$
to be nonnegative.  It can be
expressed via the present-day value $w_{\fii 0}$ of the scalar field
effective barotropic parameter \bear{}
w_{\fii}=\frac{p_{\fii}}{\rho_{\fii}}=
\frac{\dot{\fii}^2/2-V(\fii)}{\dot{\fii}^2/2+V(\fii)}.\nn
\ear
One easily obtain
$(1-q_0)/3-\Omega_{\rho 0}/2=(1+w_{\fii 0})(1-\Omega_{\rho 0})/2\geq 0$.
Therefore the model is consistent with the
observational data if the present-day value of the scalar field effective
barotropic parameter is not less than $-1$.

\begin{figure}
 \rotatebox{-90}{\epsfxsize=5cm \epsfbox{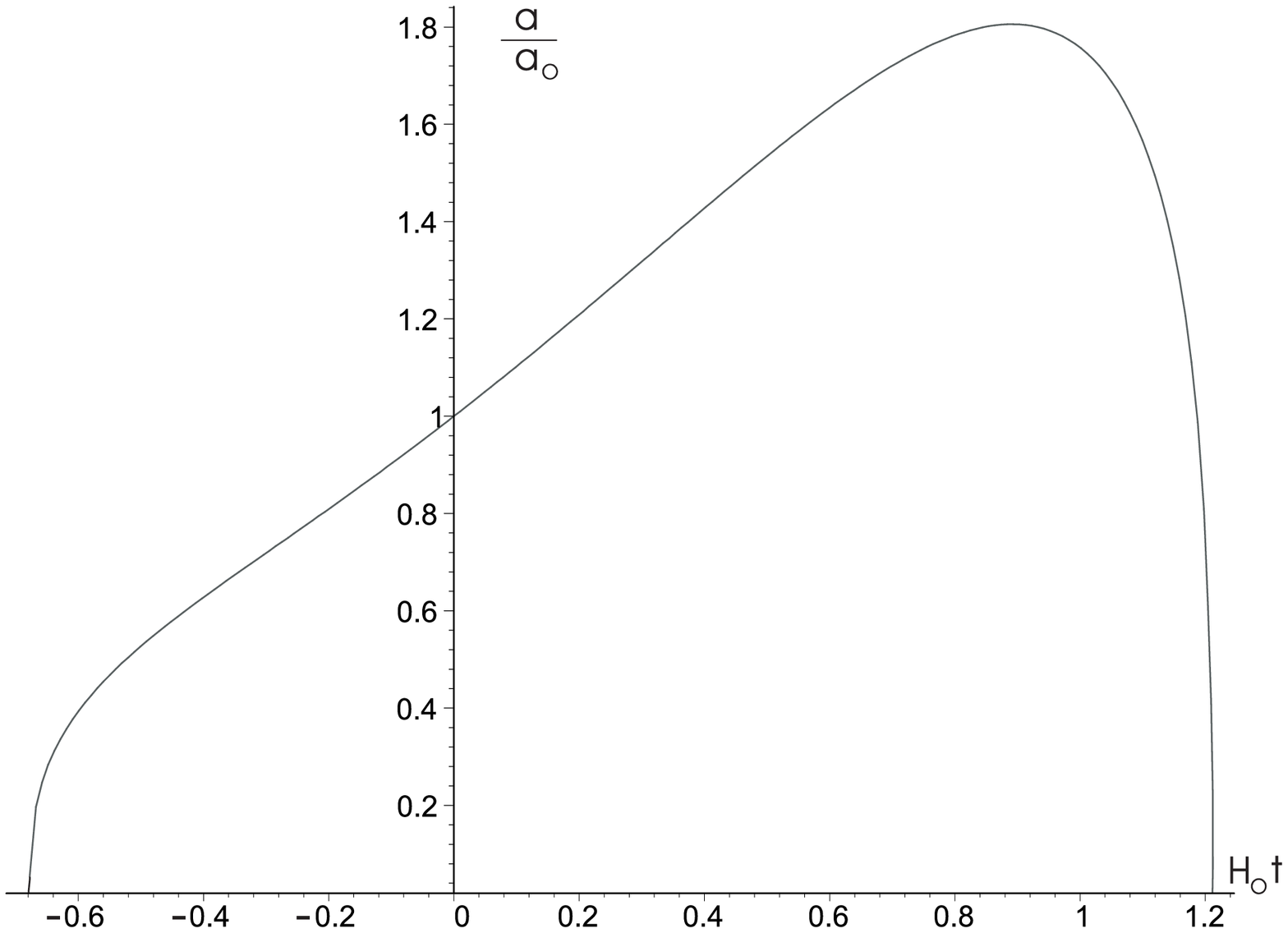}}
\hfill \rotatebox{-90}{\epsfxsize=5cm \epsfbox{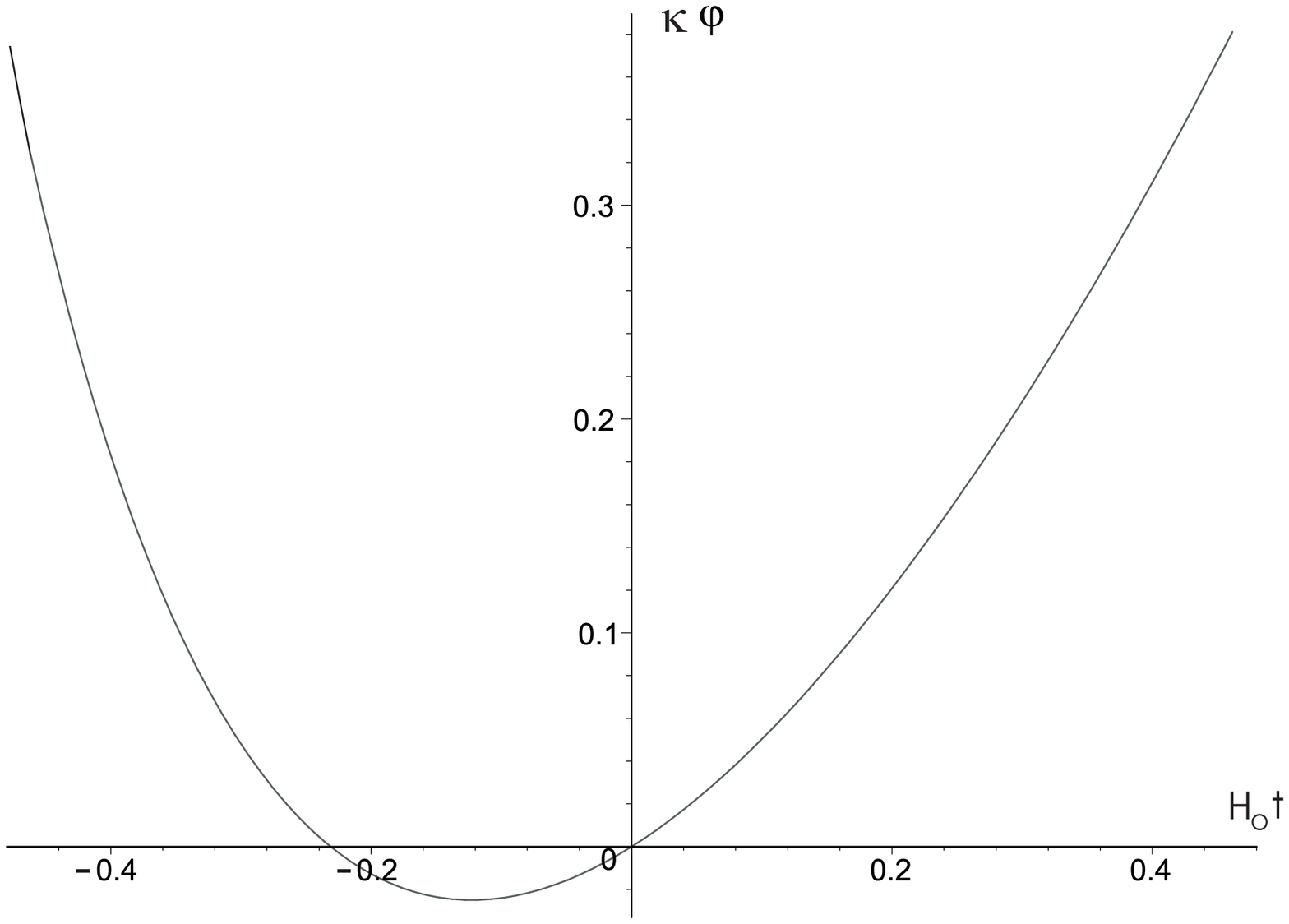}}
\\
\vskip5mm
\parbox[t]{8cm}{\parindent2cm {\bf Figure 2a}}
\hfill
\parbox[t]{8cm}{\parindent2.5cm {\bf Figure 2b}}
\\
 \rotatebox{-90}{\epsfxsize=5cm \epsfbox{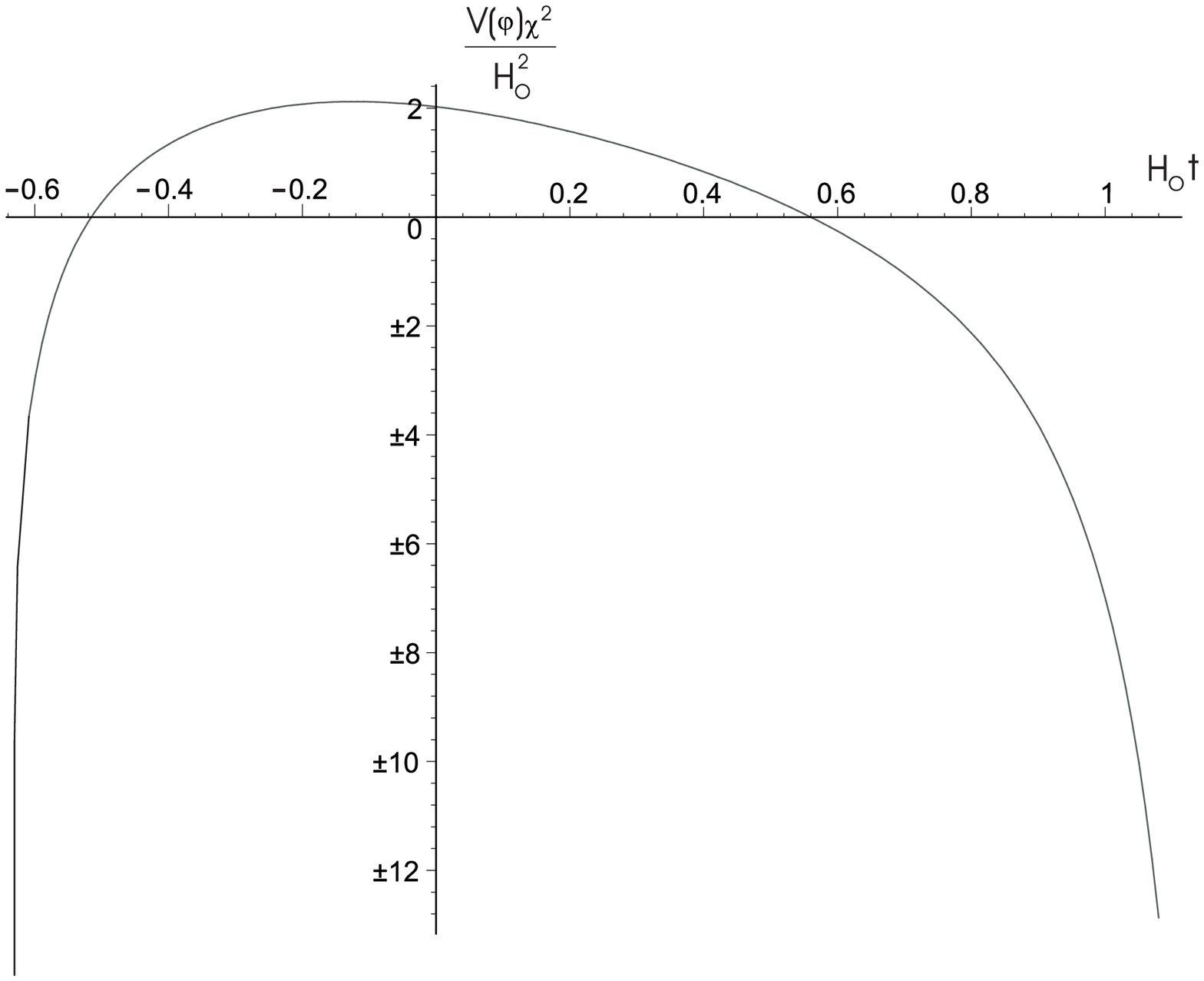}}
\hfill
 \rotatebox{-90}{\epsfxsize=5cm \epsfbox{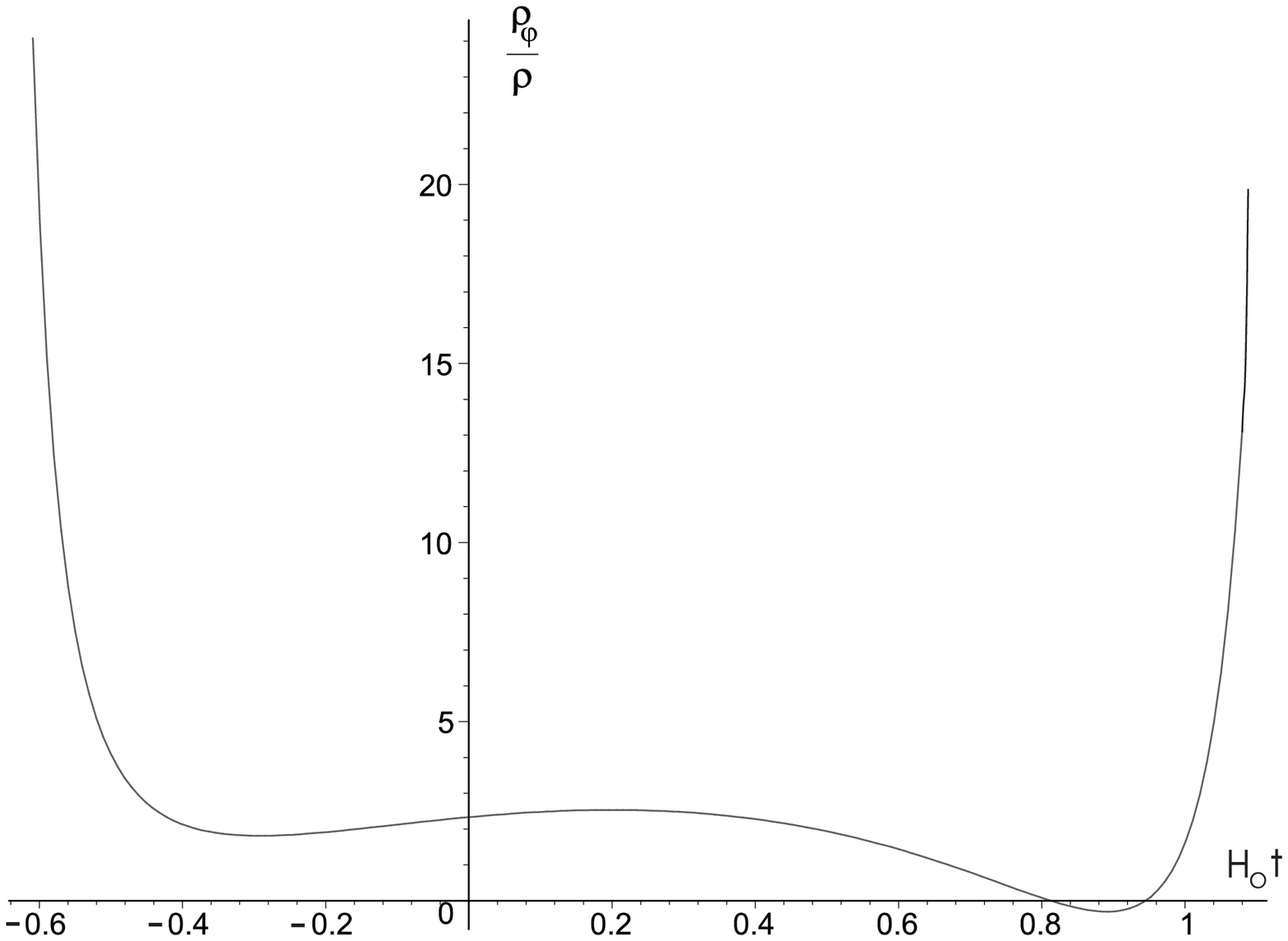}}
\\
\vskip5mm
\parbox[t]{8cm}{\parindent2cm {\bf Figure 2c}}
\hfill
\parbox[t]{8cm}{\parindent2.5cm {\bf Figure 2d}}
\vskip1cm
\parbox[c]{14cm}
{\caption{We present the exact solution for the
following  observational parameters: $q_0=0.5$, $\Omega_{\rho
0}=0.3$  and the following model parameters: $\pi
H_0/(2\om_2)=2$, $\fii_0=0$, $\dot{\varphi_0}>0$.}}
\end{figure}

The solution presented by Eqs.(\ref{13})-(\ref{17}) with the
scale factor $a$ and the scalar field $\fii$ obtained from
Eq.(\ref{7}) exists in some finite time interval
$(t_{01},t_{02})$ with negative $t_{01}$ and positive $t_{02}$
where $x>0$ and $y>0$.  Obviously, for given
observational parameters $H_0$, $q_0$ and $\Omega_{\rho 0}$ the
values $t_{01}$ and $t_{02}$ and, consequently, the full time
$(t_{02}-t_{01})$ of the universe evolution depends on
parameters $\om_1$ and $\om_2$ determining the potential of the
model. As we already mentioned in section 3, the typical time of
evolution may be approximately estimated by the value
$\pi/(2\om_2)$. Then, instead of $\om_1$ and $\om_2$ we use
further the following dimensionless model parameters:
the typical time $\pi H_0/(2\om_2)$ of the evolution in units
of $H_0^{-1}\approx 14$ billions years and
the present-day
value
$\kappa\fii(0)\equiv\kappa\fii_0=
\sqrt{2/3}\log(\sqrt{1+\varepsilon^2_0}-\varepsilon_0)$
of the scalar field $\fii$ in units of $M_p=\kappa^{-1}$.
Evidently, the model parameter $\om_1$ may be found for given
$\om_2$ and $\fii_0$ from Eq.(\ref{18}).
On Fig. 2 we present the exact solution for the following
observational parameters:
$q_0=0.5$, $\Omega_{\rho 0}=0.3$, $\dot{\varphi_0}>0$
and the following model parameters:
$\pi H_0/(2\om_2)=2$,
$\fii_0=0$.
The interval of definition of the solution is turned to be
with the following end-points
$t_{01}\approx -0.6577H_0^{-1}$
and $t_{02}\approx 1.2222H_0^{-1}$.
The scale factor presented on Fig. 2a
may be described in the main order by
$a\sim (t-t_{01})^{1/3}$
and $a\sim (t-t_{02})^{1/3}$
near the initial and final points.
The scalar
field presented on Fig. 2b diverges to $+\infty$ as
$t\to t_{01}$
or $t\to t_{02}$.
The scalar field potential given by Eq.(\ref{6})
diverges to
$-\infty$ as $t\to t_{01}$ or $t\to t_{02}$
(see Fig. 2c).
Fig. 2d shows the domination of the scalar field near the initial and
final singularities. Comparing Figs. 2c,2d one easily concludes that the
kinetic term $\dot{\fii}^2/2$ dominates the potential $V(\fii)$ near these
points. Then
$w_{\fii}\to 1$ as $t\to t_{01}$ or $t\to t_{02}$,
i.e. the
scalar field is like the stiff matter.
\begin{figure}
\begin{center}
 {\epsfxsize=12cm \epsfbox{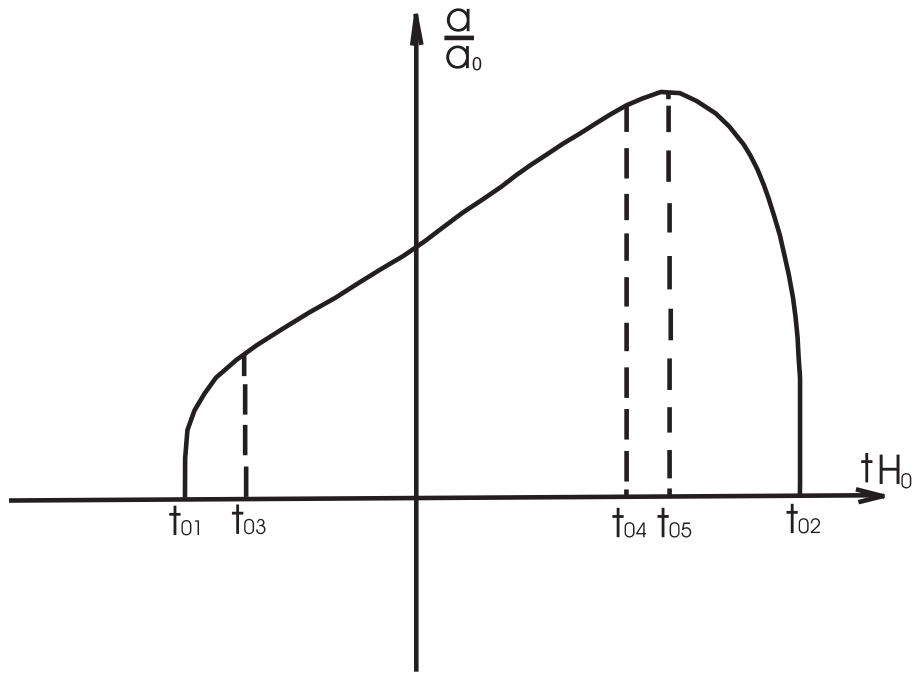}}
\end{center}
\parbox[c]{14cm}
{\caption{We introduce the following moments of time: $t_{01}$ -- big bang,
$t_{02}$ -- big crunch, $t_{03}$ -- the beginning of the acceleration,
$t_{04}$ -- the end of the accelerated stage of evolution,
$t_{05}$ -- the beginning
of the recollapsing.}}
\end{figure}

\begin{table}
\begin{center}
\begin{tabular}{|c|c|c|c|c|c|}
  \hline

  $\varphi_{0}\kappa$ & $H_0t_{01}$ & $H_0t_{03}$ & $H_0t_{04}$ & $H_0t_{05}$ & $H_0t_{02}$ \\
  \hline
-3   & -0,6872 & -0,2797 & 0,5645 & 2,1328 & 3,2765 \\
-2,5 & -0,6896 & -0,2797 & 0,5645 & 2,1107 & 3,2054 \\
-2   & -0,6896 & -0,2822 & 0,5718 & 2,0493 & 3,0213 \\
-1,5 & -0,6945 & -0,2871 & 0,5865 & 1,9021 & 2,6728 \\
-1   & -0,7019 & -0,2945 & 0,5988 & 1,6493 & 2,2162 \\
-0,5 & -0,7044 & -0,2994 & 0,5301 & 1,3057 & 1,7229 \\
0    & -0,6577 & -0,2675 & 0,3411 & 0,9056 & 1,2222 \\
0,5  & -0,5276 & -0,1767 & 0,1546 & 0,5399 & 0,7780 \\
1    & -0,3607 & -0,0834 & 0,0539 & 0,2847 & 0,4516 \\
1,5  & -0,2184 & -0,0294 & 0,0171 & 0,1374 & 0,2503 \\
2    & -0,1251 & -0,0073 & 0,0049 & 0,0638 & 0,1349 \\
2,5  & -0,0687 & -0,0024 & not found & 0,0269 & 0,0711 \\
3    & -0,0368 & not found & not found & 0,0122 & 0,0368 \\
  \hline
\end{tabular}
\end{center}
\parbox[c]{14cm}
{\caption
 {The dependence of the time values $t_{01}$, $t_{02}$,
$t_{03}$, $t_{04}$
and $t_{05}$ explained on Fig. 3 on the present-day value
$\varphi_0$ of the scalar field.}}

\end{table}

Now we analyze the behaviour of the model for various present-day
values $\varphi_{0}$ of the scalar field. Besides,
 $t_{01}$ and
$t_{02}$, we introduce the following important moments of
time:

$t_{03}$ -- the beginning of the acceleration,

$t_{04}$ --the end of the accelerated stage of evolution,

 $t_{05}$ -- the
beginning of the recollapsing (see Fig. 3).

\noindent Using the exact
solution given by Eqs. (13)-(17) with the scale factor and the
scalar field obtained from Eq. (7) for the parameter
$q_{0}=0.5$, $\Omega_{\rho 0}=0.3$ ($\dot{\varphi_0}>0$)
and various
$\varphi_0$, we numerically calculate the mentioned time values.
The result given in Table 1 shows the following property: the
interval of the accelerated expansion as well as the full interval
of evolution is shorter if the present-day value of the scalar
field is greater. The corresponding behavior of the scalar factor
is presented on Fig. 4.

\begin{figure}
\begin{center}
 {\epsfxsize=12cm \epsfbox{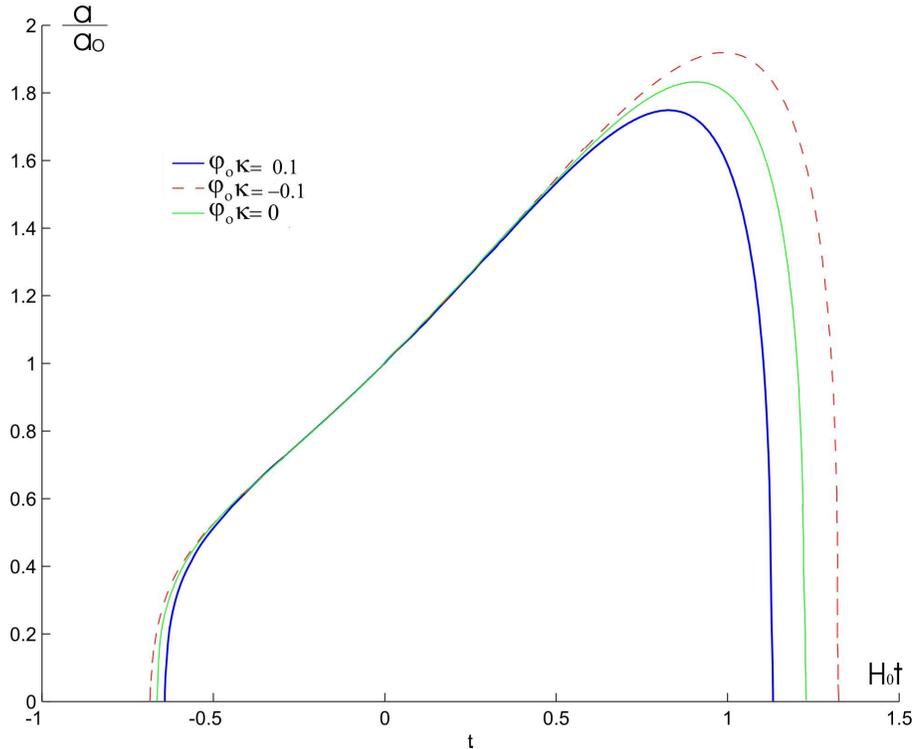}}
\end{center}
\parbox[c]{14cm}
{\caption{The scale factor for various present-day values
                  $\varphi_0$ of the scalar field. }}
\end{figure}

\section{Conclusions}
We studied the spatially flat Friedmann model containing a
pressureless perfect fluid (dust) and a minimally coupled scalar
field with an unbounded from below potential of the form
 $V(\fii)=W_0-V_0\sinh\left(\lambda\sqrt{8\pi
G_N}\fii\right)$, where the parameters $W_0$ and $V_0$ are
arbitrary. The parameter $\lambda$ is chosen to be $\sqrt{3/2}$ in
order to integrate the model in the explicit form. All exact
solutions describe the recollapsing universe. The behavior of the
model near both initial and final points of evolution is analyzed.
Near the singularity the scale factor $a\sim t^{1/3}$ for almost
all solutions. In this case the constraint Eq.(\ref{1}) is
dominated by the kinetic energy of the scalar field. Such
solutions are called kinetic dominated \cite{Wands}. There exists
a special solution with $a\sim t^{2/3}$ near the singularity. It
appears when the constraint Eq.(\ref{1}) is dominated by the
density of dust. It's so called fluid dominated solution
\cite{Wands}. The scalar field typically diverges to $\pm\infty$
as the time tends to the initial or final values. However, there
is a special solution where the scalar field tends to some
constant.

The evolution between big bang and big crunch crucially depends
on the model parameters $\om_1$ and $\om_2$. In particular, the
time of evolution is determined by these parameters. We show
that the model may be consistent with the observational parameters.
We singled out the exact solution with the present-day values of
the acceleration parameter $q_0=0.5$ and the dark matter density
parameter $\Omega_{\rho 0}=0.3$  describing the evolution within the time
approximately equal to $2H_0^{-1}$.

\ack

 This work was supported in part by the Russian Foundation
for Basic Research (Grant 01-02-17312) and DFG Project 436 RUS
113/678/0-1.We are grateful to professor A.A. Starobinsky for
helpful discussions. V.N.M. thanks Prof.Dr. Heinz Dehnen for the hospitality
at the University of Konstanz during his stay there in October-November, 2003.

\pagebreak


\Bibliography{99}
\bibitem{Mel1}
V.~N. Melnikov, in: {\it   Gravitation and Cosmology},
Proc. Int. Conf. on Gravitation and Cosmology (Ed. M. Novello), Rio de
Janeiro 1993, (Singapore: Edition Frontieres 1994) p. 147.

\bibitem{Mel2}
V.~N. Melnikov,  {\it  Multidimensional   Classical   and   Quantum
Cosmology and Gravitation:  Exact Solutions and Variations  of  Constants},
CBPF-Notas de Fisica-051/93, Rio de Janeiro 1993, 93 pp.

\bibitem{Mel3}
V.~N. Melnikov, in: {\it Gravitation and Cosmology II}
(Ed. M. Novello), (Singapore: Edition Frontieres 1996), p. 465.

\bibitem{Mel4}
V.~N. Melnikov, {\it Multidimensional Cosmology and Gravitation},
CBPF-MO-002/95,  Rio de Janeiro 1995, 210 pp.

\bibitem{Mel5}
V.~N. Melnikov, {\it Exact Solutions in Multidimensional Gravity and Cosmology III},
CBPF-MO-003/02, Rio de Janeiro 2002.


\bibitem{FJ}
P.~G. Ferreira  and M. Joice,
{\it Physical Review} {\bf D 58} 023503 (1998),
astro-ph/9711102.

\bibitem{T}
P.~K. Townsend, {\it Quintessence from M-theory}, hep-th/0110072.

\bibitem{star}
A.~A. Starobinsky, {\it JETP Lett.} {\bf 68} 757 (1998).

\bibitem{Wands}
I.~P.~C. Heard and D. Wands, {\it Cosmology with positive and
negative exponential potentials}, gr-qc/0206085.

\bibitem{Dehnen}
H. Dehnen, V.~R. Gavrilov and Melnikov,
{\it Grav. Cosm.} No. 3 (35) 189 (2003), gr-qc/0212107.

\bibitem{Linde}
R. Kallosh and A. Linde,
{\it Dark Energy and the Fate of the Universe},  astro-ph/0301087.

\endbib
\end{document}